\documentclass[journal]{IEEEtran}

\ifCLASSINFOpdf
\else
\fi

\usepackage[dvipsnames]{xcolor}
\usepackage[cmex10]{amsmath}
\usepackage{amssymb}   
\usepackage{amsxtra}
\usepackage{amscd}
\usepackage{amsthm}
\usepackage{textcomp}
\usepackage{graphicx}  
\usepackage{mathtools}
\usepackage{stfloats}
\usepackage{blindtext}
\usepackage{bigints}
\usepackage{color,soul}
\usepackage{array}  

\usepackage{cite}
\usepackage{multirow}

\begin{document}
%
\title{{\LARGE Modeling and Analysis of Reconfigurable Intelligent Surfaces for Indoor and Outdoor Applications in Future Wireless Networks}}

%
%
%

\author{Ibrahim~Yildirim,~\IEEEmembership{Student~Member,~IEEE,} Ali~Uyrus,~\IEEEmembership{Student~Member,~IEEE} and~ Ertugrul~Basar,~\IEEEmembership{Senior~Member,~IEEE} 
\thanks{I. Yildirim is with Department of Electronics and Communication Engineering, Istanbul Technical University, Istanbul, Turkey. e-mail: yildirimib@itu.edu.tr}
\thanks{A. Uyrus and E. Basar are with the Communications Research and Innovation Laboratory (CoreLab), Department of Electrical and Electronics Engineering, Koç University, Sariyer 34450, Istanbul, Turkey. e-mail: auyrus18@ku.edu.tr and ebasar@ku.edu.tr}
\thanks{The work of E. Basar was supported by the Scientific and Technological Research Council of Turkey (TUBITAK) under Grant 117E869 and the Turkish Academy of Sciences (TUBA) GEBIP Programme}
}

%
%

\markboth{}%
{}
%



\maketitle
\begin{abstract}
Reconfigurable intelligent surface (RIS)-empowered communication is one of the promising 6G technologies that allows the conversion of the wireless channel into an intelligent transmit entity by manipulating the impinging waves using man-made surfaces.
In this paper, the potential benefits of using RISs are investigated for indoor/outdoor setups and various frequency bands (from sub $ 6 $ GHz to millimeter-waves). First, a general system model with a single RIS is considered and the effect of the total number of reflecting elements on the probabilistic distribution of the received signal-to-noise ratio and error performance is investigated under Rician fading. Also for this case, the path loss exponent is analyzed by considering empirical path loss models.
Furthermore, transmission models with multiple RISs are developed and analyzed for indoor and outdoor non line-of-sight (NLOS) scenarios. The conventional RIS selection strategies are also integrated for systems equipped with multiple RISs for the first time.  
Through extensive simulations, it is demonstrated that the RIS-assisted systems provide promising solutions for indoor/outdoor scenarios at various operating frequencies and exhibit significant results in error performance and achievable data rates even in the presence of system imperfections such as limited range phase adjustment and imperfect channel phase estimation at RISs.  

\end{abstract}


%
\IEEEpeerreviewmaketitle

\section{Introduction}
\IEEEPARstart{O}{ne} of the areas that has made the most progress as a result of the digitalization experienced by human life in recent decades is wireless communications \cite{6G_roadmap}.
Due to the enormous increase in data traffic, existing communication systems have faced challenges in providing high quality of service. Both academia and industry have carried out intensive research activities in order to overcome the deficiencies of legacy transmission concepts. All these efforts and targets have attracted the attention of the wireless community in recent times, most notably within the context of fifth-generation (5G) wireless networks. 3rd Generation Partnership Project (3GPP) Release 15, which is the first full set of 5G standard, has been completed in 2018. This initial 5G standard enables high data rate, ultra-reliability and low latency using novel technologies, such as scalable orthogonal frequency-division multiplexing (OFDM) numerologies, millimeter-wave (mmWave) wireless communications and massive multiple-input multiple-output (MIMO) systems \cite{Ericsson}. Massive machine type communications, enhanced mobile broadband, and ultra-reliable and low latency communications are defined as three use-cases with diverse requirements and applications in 5G wireless networks. Beyond all these approaches, more radical physical layer concepts are needed to comprehend the potential requirements of future networks due to sophisticated applications with extremely high quality-of-service (QoS) requirements, such as augmented reality, massive Internet-of-everything, and autonomous vehicles. Despite the high expectations, 5G has not brought these sophisticated applications into reality to date \cite{6G_Rappaport}. This has pushed researchers to look new paradigms beyond 5G and start to conceptualize sixth-generation (6G) wireless networks. Within the frame of 6G studies, it is aimed to design systems that are expected to have substantial differences from previous generations by achieving a radical transformation in wireless networks.

In light of the above, it is also essential to develop flexible, versatile, and inclusive physical layer solutions in order to support potential requirements of 6G wireless networks. Although researchers have put forward promising solutions for beyond 5G, including index modulation schemes \cite{IM_5G}, non-orthogonal multiple access \cite{NOMA}, and alternative waveforms \cite{Waveform_2014}, the overall progress has been still relatively slow in terms of QoS, reliability and security.

In radio communication environments, one of the most deteriorating characteristics of the channel is its random and dynamic nature. However, when the spectrum shortage below $ 6 $ GHz frequency bands has taken into account, it is inevitable to migrate to the mmWave ($ 30 $-$ 100 $ GHz) and THz (above $ 100 $ GHz) bands. In above $ 6 $ GHz, the signal will be more susceptible to blockage and interference, since collecting high energy in the receiver will be difficult because of small antenna sizes. Therefore, the severe signal attenuation and blockage prevent mmWaves to reach long distances. In order to ensure reliable wireless transmission and to alleviate these disruptive effects at high frequencies, we need to overcome the negative and uncontrollable effects of the wireless channel. 

Most of the promising methods for next-generation wireless networks are mainly based on the intelligence or reconfigurability of the building blocks of communication systems. Recently, reconfigurable intelligent surfaces (RISs) have been put forward by researchers to enable the control of wireless environments via their unique and effective functionalities, such as wave absorption, anomalous reflection, polarized reflection, wave splitting, wave focusing, and phase modification \cite{Tan_2016,Tan_2018,9136592}. Recent results have revealed that these attractive electromagnetic functionalities are possible without complex operations such as decoding, encoding, and radio frequency (RF) processing and the communication system performance can be enhanced by exploiting the implicit randomness of wireless propagation \cite{Di_Renzo_2019,Basar_2020_TCOM,Basar_Access_2019,Huang_2019}.

The use of RISs is particularly useful when the line-of-sight (LOS) link is blocked or not strong enough since it is possible to provide additional transmission paths by utilizing the reflecting elements of an RIS. Although it is possible to alleviate the negative effects of the channel using relays, the hardware cost, power consumption and latency increase considerably as the signal has been actively processed at each relay \cite{Ntontin_relay}. In addition, real-time controlled phase shifts of each RIS element allow the optimization of certain system performance metrics, such as transmit power, achievable rate, energy efficiency, and received signal-to-noise ratio (SNR) \cite{Wu_2018, Wu_2018_2,Huang_2018_2}. On the other hand, various secrecy enhancing schemes have been proposed by optimizing the transmit beamformer and RIS phase shifts jointly in \cite{Cui_2019,Schober_2019_2, Shen_2019}. In \cite{Basar_2019_LIS}, the communication through RISs has been investigated in terms of error performance and the symbol error probability (SEP) is derived for a scenario without a direct path between the transmitter and the receiver. However, to the best of our knowledge, a general analysis that elaborates the performance of single and multiple RIS-assisted systems in indoor and outdoor scenarios as well as under different conditions is missing in the literature.

In this paper, we investigate the performance of single and multiple RIS-assisted systems without a direct path between the transmitter and the receiver in indoor and outdoor propagation environments. First, in Section II, we show that the employment of an RIS is a dominant factor in the received signal power even in the absence of a direct communication link. By deriving mathematical expressions for the error performance under both large-scale and small-scale fading effects, we show that the RIS-assisted link acts as a superior LOS path by suppressing the the disruptive effects of the propagation environment. Furthermore,  we investigate the effects of RISs on the path loss exponent (PLE) for the log-distance path loss model \cite{Rappaport}, and determine the achievable data rate of an RIS-assisted system under emprical path loss models for below and above $ 6 $ GHz operating frequencies. 
In Section III, we investigate the wireless communications in the presence of multiple RISs under two envisioned scenarios: simultaneous transmission over two RISs in indoor environment and the double-RIS reflected transmission in outdoor environment. We introduce a mathematical framework on the error performance for these two scenarios and demonstrate that the use of multiple RISs can be a remedy for the blockage problem in 6G and beyond systems. We also introduce RIS selection strategies for systems equipped with multiple RISs. Finally, we study the RIS imperfections, such as practical phase shifts and phase estimation errors, and demonstrate the robustness of RIS-assisted systems. Our comprehensive theoretical and numerical results in Section IV show that the transmission through RISs appear as an attractive candidate for indoor/outdoor communication systems at various operating frequencies by suppressing the destructive effects of wireless propagation. 



\begin{figure}[!t] 
	\begin{center}\resizebox*{7.5cm}{5.5cm}{\includegraphics{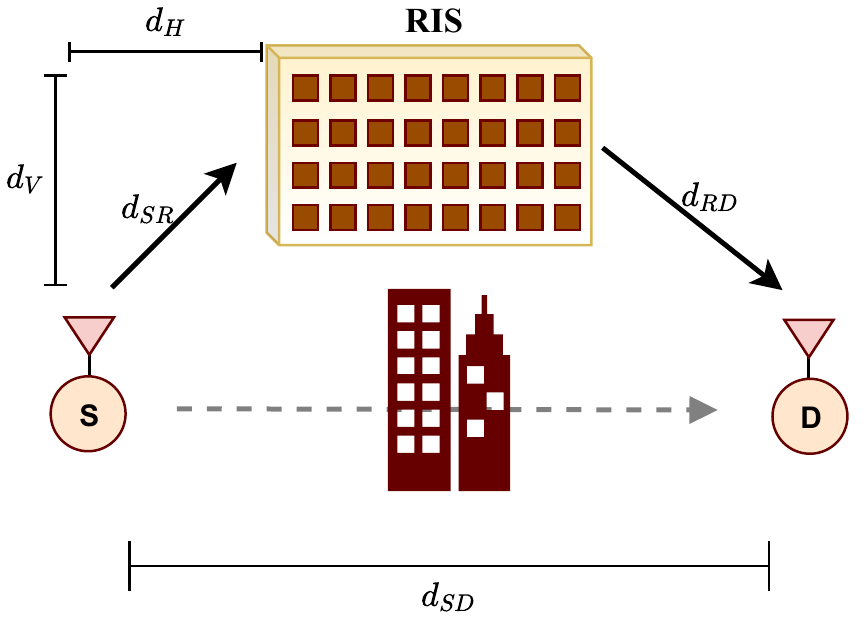}}
		\vspace*{-0.0cm}\caption{An RIS-assisted SISO system consisting of a direct path and an RIS with $N$ reflecting elements.}\vspace*{-0.5cm}\label{single_RIS_model}
	\end{center} 
\end{figure}

\section{Communications Through a Single RIS}
In this section, we consider the wireless communication system shown in Fig. \ref{single_RIS_model} with a transmitter/receiver pair and a single RIS.
Here, we provide first time a unified error  performance analysis that is applicable to various path loss 
models in indoor and outdoor communication scenarios. Furthermore, the effect of an RIS on the overall path loss is examined in frequency bands below and above $6$ GHz. 

\subsection{The General Model and Performance Analysis}
As illustrated in Fig. \ref{single_RIS_model}, a generic RIS-assisted communication scenario is considered, where the transmission is carried out via an RIS under the blocked direct link between the source (S) and the destination (D).
We assume that the real-time controlled RIS 
is equipped with $ N $ reconfigurable reflect elements that can be adjusted according to channel phases. In Fig. \ref{single_RIS_model} and the remainder of the paper we use the notations  $d_{SD}$, $d_{SR}$ and $d_{RD}$ to represent the distances from source-to-destination (S-D), source-to-RIS (S-R), and RIS-to-destination (R-D), respectively. In this setup, $d_V$ and $d_H$ respectively denote the vertical and horizontal distance between S and RIS. Under the Rician fading assumption, the channel fading coefficients of S-R and R-D links for $i$th reflecting element ($i=1, \ldots,N$) are respectively represented by
\begin{align} \label{}
h_i^{SR}=& \sqrt{\frac{K_1}{1+K_1}}\bar{h}_i^{SR}+\sqrt{\frac{1}{1+K_1}}\tilde{h}_i^{SR}, \nonumber \\
h_i^{RD}=& \sqrt{\frac{K_2}{1+K_2}}\bar{h}_i^{RD}+\sqrt{\frac{1}{1+K_2}}\tilde{h}_i^{RD}
\end{align}
where $K_1$ and $K_2$ are the Rician factors of these two links, $\bar{h}_i^{SR}$ and $\bar{h}_i^{RD}$ stand for the LOS components, and $\tilde{h}_i^{SR}$ and $\tilde{h}_i^{RD}$ represent the NLOS components for $i=1, \ldots,N$. Here, $\tilde{h}_i^{SR}$ and $\tilde{h}_i^{RD}$  follow $\mathcal{CN}(0,1) $ distribution.  

Under the condition of slow and flat fading channels, the baseband received signal at D can be expressed as 	\begin{align} \label{receive_sig}
r=\sqrt{\it{p_t}}\bigg[\sqrt{P_{L}^R}\left( \sum\nolimits_{i=1}^{N}{ h_i^{SR}e^{j\phi_i}h_i^{RD}} \right)\bigg]x+w
\end{align}
where $p_t$ is the total transmitted power, $\phi_i$ stands for the controllable phase shift prompted by the $i$th RIS element under the assumption of unit-gain reflection coefficients, $x$ is the data symbol selected from $M$-ary phase shift keying/quadrature amplitude modulation (PSK/QAM) constellations with unit power,  and $w \sim \mathcal{CN}(0,N_0) $ represents the additive white Gaussian noise (AWGN) sample, where $N_0$ is the total noise power. Additionally, $P_L^R$ stands for the path loss of the RIS-assisted path. Under the specular reflection in the near-field case \cite{Fontan_2008}, $P_L^R$ will be proportional to total length of the RIS-assisted path \small$(d_{SR}+ d_{RD})^2$\normalsize, while $P_L^R$ will be a proportional with $d_{SR}^2d_{RD}^2$ under plate scattering paradigm in far-field case \cite{Ellingson}. All analyses and derivations in this work are performed by considering reflect-array type RISs in which the RIS elements are separated with half-wavelength under the far-field case. 
Using well known radar range equation, the path loss $P_L^R$ can be expressed as
\begin{align}\label{ellingston_PL}
P_L^R =  \left(\left(\frac{\lambda}{4\pi}\right)^4\frac{G_e^iG_e^r}{d_{SR}^2d_{RD}^2} \epsilon_p\right)^{-1}
\end{align}
where $\lambda$ is the wavelength, $G_e^i$ is the gain of the RIS in the direction of incoming wave, $G_e^r$ is the gain of RIS in the direction of received wave and $\epsilon_p$ is the efficiency of RIS, which is described as ratio of power transmit signal power by RIS to received signal power by RIS. In this paper, it is assumed that that RIS consist of passive elements and $\epsilon_p=1$. 

Expressing the channel fading coefficients as  ${h_i^{SR}}=\alpha_ie^{-j\theta_i}$ and ${h_i^{RD}}=\beta_ie^{-j\varphi_i}$, where $\alpha_i$ and $\beta_i$ stand for channel amplitudes while $\theta_i$ and $\varphi_i$ denote channel phases, the received signal can be rewritten as \vspace*{-0.2cm}
\begin{align}\label{single_RIS}
r = \sqrt{\it{p_t}}\bigg[\sqrt{P_{L}^R}\left( \sum\nolimits_{i=1}^{N}{\alpha _i\beta_ie^{j\Delta \Phi_i}} \right) \bigg]x+w
\end{align}
where $\Delta \Phi_i=\phi_i-\theta_i-\varphi_i$ is the phase difference term for $i=1,...,N$. Therefore, the instantaneous SNR at D is given by
\begin{align} \label{SNR_1}
\gamma \!=\!\frac{{{\left|\sqrt{P_{L}^R}\left( \sum\nolimits_{i=1}^{N}{\alpha_i\beta _ie^{j\Delta \Phi_i}} \right) \right|}^2}{\it{p_t}}}{N_0}.
\end{align}
Here, the SNR can be maximized by aligning the phases of the reflected signals from the RIS to the phase of the direct path, that is, $\Delta \Phi_i=0 $ for $i=1,2,...,N$ considering trigonometric identities \cite{Basar_2019_LIS}.
Under this intelligent reflection model with manipulated reflection phases, the maximized instantaneous SNR at D is obtained as \vspace*{-0.1cm}
\begin{align}\label{SNR1_max}\small
\gamma_{\max} =\frac{{{\left(  \sqrt{P_{L}^R}\left( \sum\nolimits_{i=1}^{N}\alpha _i\beta_i \right) \right)}^2}{\it{p_t}}}{N_0}=\frac{{A^2}{{\it{p_t}}}}{N_0}
\end{align}\normalsize
where $\alpha_i$ and $\beta_i$ follow Rician distribution. Using the central limit theorem (CLT), it can be shown that $A$, which is the sum of $N$ independently identical distributed (iid) random variables, follows the Gaussian distribution for $N\gg1$ with the following mean and variance:
\begin{align} \small
\text{E}[A]=&N\frac{\sqrt{P_L^R}\pi }{4(K+1)}\bigg( L_{1/2}\left(-\frac{K^2}{(K+1)} \right) \bigg)^2, \nonumber \\
\text{VAR}[A] =&NP_L^R -N \frac{{P_L^R}\pi^2 }{16(K+1)^2}\bigg( L_{1/2}\left(-\frac{K^2}{(K+1)} \right) \bigg)^4
\end{align} \normalsize
where $L_{1/2}(x)$ is the Laguerre polynomials of degree $1/2$.
Since $\gamma$ follows non-central chi-square distribution with one degree of freedom, its moment generating function (MGF) \cite{Proakis} is obtained as
\begin{align}
M_{\gamma_{\max}}(s)=  \frac{\exp\left( \frac{\frac{sN^2P_L^R\pi^2L_{1/2}^4\left(-K^2/(K+1)\right) \it{p_t}}{16(K+1)^2N_0}}{1-\frac{sNP_L^R\left[ 16(K+1)^2-\pi^2L_{1/2}^4\left(-K^2/(K+1)\right)\right]\it{p_t}}{8(K+1)^2N_0}} \right)  }{\sqrt{1-\frac{sNP_L^R\left[ 16(K+1)^2-\pi^2L_{1/2}^4\left(-K^2/(K+1)\right)\right]\it{p_t}}{8(K+1)^2N_0}}}.
\end{align}

Considering the generic SEP expression of \cite{Simon} for  $M$-PSK signaling, we obtain
\begin{align}
P_e = \frac{1}{\pi}\int_0^{(M-1)\pi/M} M_{\gamma_{\max}} \left(-\frac{\sin(\pi/M)}{\sin^2\eta}\right)d\eta.
\end{align}
When binary PSK (BPSK) signaling is used, the average SEP can be simplified as	
\begin{align} \label{bep_bpsk1}
&P_e = \frac{1}{\pi} {\bigintsss_0^{\pi/2}} \frac{1}{\sqrt{1+\frac{NP_L^R\left[ 16(K+1)^2-\pi^2L_{1/2}^4\left(-K^2/(K+1)\right)\right]\it{p_t}}{8(K+1)^2N_0\sin^2\eta}}} \nonumber \\
&\times \exp\left( -\frac{\frac{N^2P_L^R\pi^2L_{1/2}^4\left(-K^2/(K+1)\right) \it{p_t}}{16(K+1)^2N_0\sin^2\eta}}{1+\frac{NP_L^R\left[ 16(K+1)^2-\pi^2L_{1/2}^4\left(-K^2/(K+1)\right)\right]\it{p_t}}{8(K+1)^2N_0\sin^2\eta}} \right)  d\eta.   
\end{align}
Here, the average SEP in (\ref{bep_bpsk1}) can be further upper bounded by letting $\eta=\pi/2$.
As will be shown in Section IV, the average SEP decreases with $N^2$ for the low $p_t/N_0$ region. Furthermore, we observe from (\ref{bep_bpsk1}) that for sufficiently large $ N $, the effect of the blocked direct path becomes less significant since the signals reflected from the RIS compensate deficiencies in the received SNR. Using the similar methods in \cite{Basar_2020_TCOM} and \cite{Basar_Access_2019}, the average SEP is derived under the Rician fading channel and the path loss is also taken into account in \eqref{bep_bpsk1}. Finally, we note that \eqref{bep_bpsk1} can be used under different path loss models and generalizes the SEP derivations of \cite{Basar_Access_2019}.

\subsection{Performance under Empirical Path Loss Models for Outdoor Applications: Below and Above $6$ GHz}
In this subsection, we examine the effect of an RIS on the overall path loss between S and D in frequency bands below and above $ 6 $ GHz for the scenario of Fig. \ref{single_RIS_model}. Empirical path loss models for outdoor communication scenarios have been taken into account to assess the potential of RIS-assisted systems.
Then, the impact of an RIS on the PLE and achievable data rate is observed under the empirical transmission scenarios.

\subsubsection{Urban Micro (UMi) Path Loss Models with a Single RIS}
\begin{figure}[t] 
	\begin{center}\resizebox*{7.5cm}{6.4cm}{\includegraphics{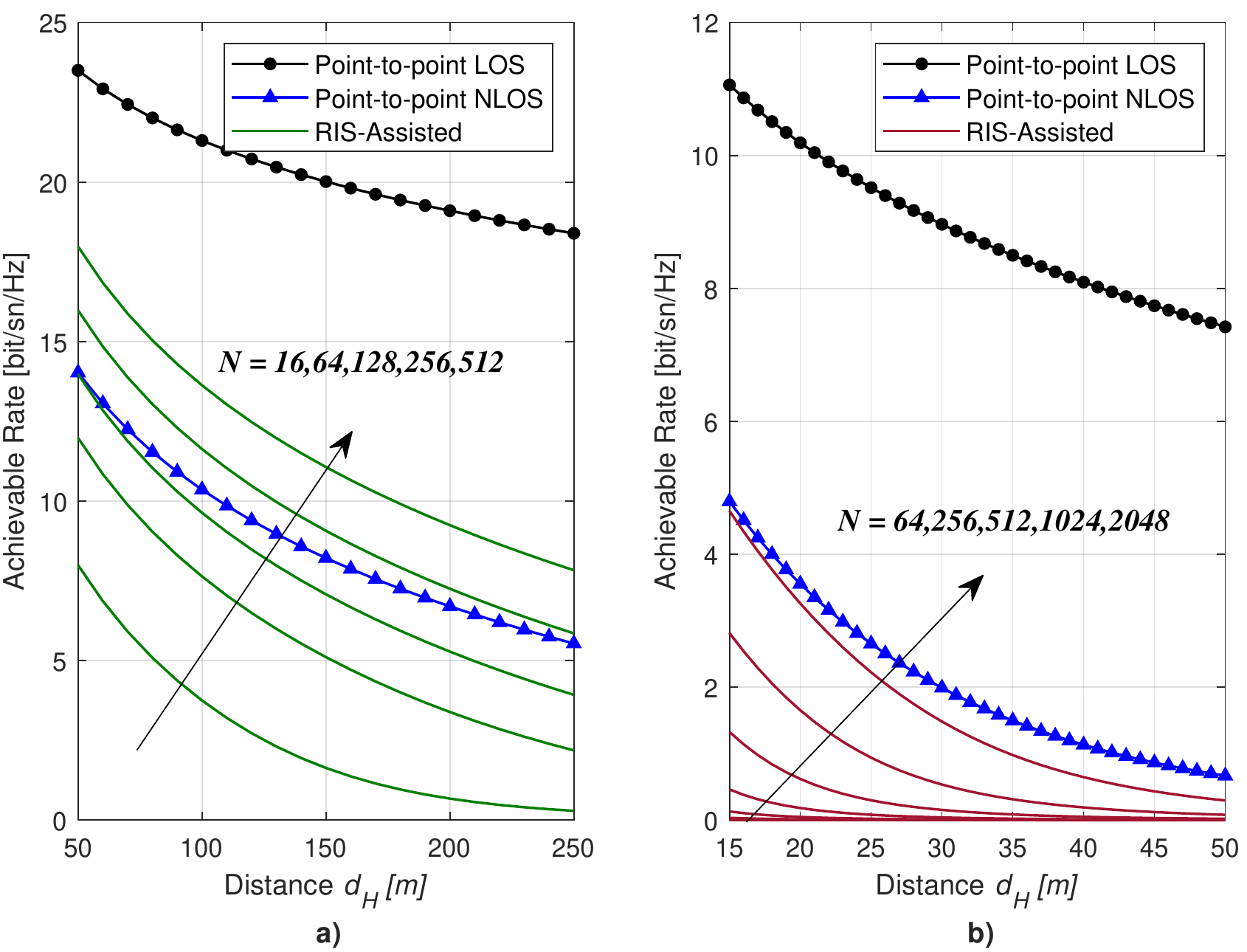}}
		\vspace*{-0.0cm}\caption{The achievable data rate comparison for direct and RIS-assisted transmission under (a) 3GPP UMi path loss model with $f_c = 2.4$ GHz and (b) 5G UMi-Street Canyon path loss model with $f_c = 28$ GHz. }\vspace*{-0.5cm}\label{fig:data_rate}
	\end{center}
\end{figure}
\normalsize
In order to assess the potential of an RIS over an empirical transmission environment, we consider the 3GPP UMi \cite{3GPP_phy2} and 5G UMi-Street Canyon \cite{5GCM_Rappaport} path loss models. As a reference, a point-to-point LOS and NLOS transmission scenarios with a single transmitter/receiver path are considered. At the $2$-$6$ GHz frequency band and S-D distance ranging from $10$ m to $2000$ m, 3GPP UMi path loss models are respectively expressed for LOS and non-LOS (NLOS) transmission as \cite{3GPP_phy2}
\begin{align}
{P}_{L}(d) \, [\text{dB}]&=22\log_{10}(d)+28+20\log_{10}(f_c),  \nonumber \\
{P}_{NL}(d)\,[\text{dB}]&=36.7\log_{10}(d)+22.7+26\log_{10}(f_c).		\end{align}

At frequency bands from $6$ GHz to $100$ GHz, the UMi-Street Canyon path loss models both LOS and NLOS cases are respectively given by \cite{5GCM_Rappaport}   
\begin{align}
{P}_{L}(d) \, [\text{dB}]&=21\log_{10}(d)+32.4+20\log_{10}(f_c),  \nonumber \\
{P}_{NL}(d)\,[\text{dB}]&=31.7\log_{10}(d)+32.4+20\log_{10}(f_c).		\end{align}
The received signals at D for RIS-assisted scenarios with $N$ reflecting elements is obtained as in \eqref{receive_sig} under plate scattering paradigm with \small${{P}^{R}_{L}}\propto {P}_L(d_{SR}){P}_L(d_{RD})$\normalsize, that is, assuming LOS links for RISs. By considering the same analysis in \eqref{SNR_1}, the achievable data rate expressions for a single RIS-assisted system is given by
\begin{equation}
R_{RIS} = \log_2 \left( 1+  \gamma_{RIS}\right)
\end{equation} 
where $\gamma_{RIS}$ is the received SNR calculated by \eqref{SNR_1}.

For the RIS-assisted scenario, the achievable data rate is maximized by aligning the phases of the reflected signals from the RIS to the phase of the direct path with  $\Delta \Phi_i=0 $ for $i=1,2,...,N$ as in \eqref{SNR1_max}.

In Fig. \ref{fig:data_rate}, achievable data rates under 3GPP and 5G UMi path loss models are respectively given with respect to varying horizontal distance ($d_H$) between S and RIS for $d_{SD}=4d_H$, $d_V=10$ m, $N_0=-95$ dBm and $p_t=5$ W. The results are obtained for an operating frequency of $2.4$ GHz for the 3GPP UMi path loss model, while $28$ GHz is considered for the 5G UMi-Street Canyon path loss model. Here, the point-to-point physical NLOS link between the S and the D is modeled with Rayleigh distribution and considering the physical NLOS path loss models in (11) and (12), while the point-to-point LOS link is modeled with Rician distribution and LOS path loss models. In both scenarios, a reliable communication is provided by eliminating the effect of the blocked direct link with increasing $N$ values. It should be also noted that the closer the RIS is placed to the source, the higher the achievable data rate to be reached. Particularly, at high frequencies, in order to transmit at the level of point-to-point LOS scenario, it is necessary to make huge increases in number of reflecting elements, while the same rates can be achieved by using fewer reflectors at $ 2.4 $ GHz. Particularly, as shown in Fig. \ref{fig:data_rate}(b), the achievable rate shows a rapid downfall with increasing distances, since the signals in the mmWave bands are more susceptible to path loss. 

\subsubsection{Analysis of Path Loss Exponent Under RISs}

Here, we investigate the variation of the average total path loss with respect to distance in an RIS-assisted communication channel and the effect of an RIS on PLE under the log-distance path loss model for different number of reflecting elements for the scenario of Fig.\ref{single_RIS_model}.

3GPP UMi and 5G UMi-Street Canyon path loss models, which are discussed in previous subsection, are utilized to express the path loss of sub-paths, which are defined from S to RIS ($P_L(d_{SR})$) and from RIS to S ($P_L(d_{RD})$). Total received power at the D through the $i$th RIS element is calculated under the plate scattering paradigm {\cite{Ellingson} as:
	
	\begin{equation} \label{rec_pow_single_RIS}
	p_{r_{i}} = \frac{p_tG_T|\Gamma_i|G_R}{P_L(d_{SR})P_L(d_{RD})}
	\end{equation}
where $G_T$ and $G_R$ are gains of the transmitter and receiver antennas, respectively. $\Gamma_i$ is the reflection coefficient of the $ i $th RIS element, which is given by
\begin{equation} \label{Gamma}
\Gamma_i = e^{-j\varphi_i} G_e^iG_e^r\epsilon_b.
\end{equation}
where $\varphi_i$ is the phase difference inducted by $i$th RIS element.

Total received power at the receiver including all RIS elements is expressed as:

\begin{equation} \label{rec_pow_all_RIS}
p_{r_{i}} = \left(\sum\nolimits_{i} \sqrt{  \frac{p_tG_T|\Gamma_i|G_R}{P_L(d_{SR})P_L(d_{RD})}}e^{j\phi_i}\right)^2
\end{equation}
where $\phi_i$ represents the phase delay of the signal received through $ i $th RIS element.

\begin{figure}[t]
	\begin{center}\resizebox*{7.5cm}{6.4cm}{\includegraphics{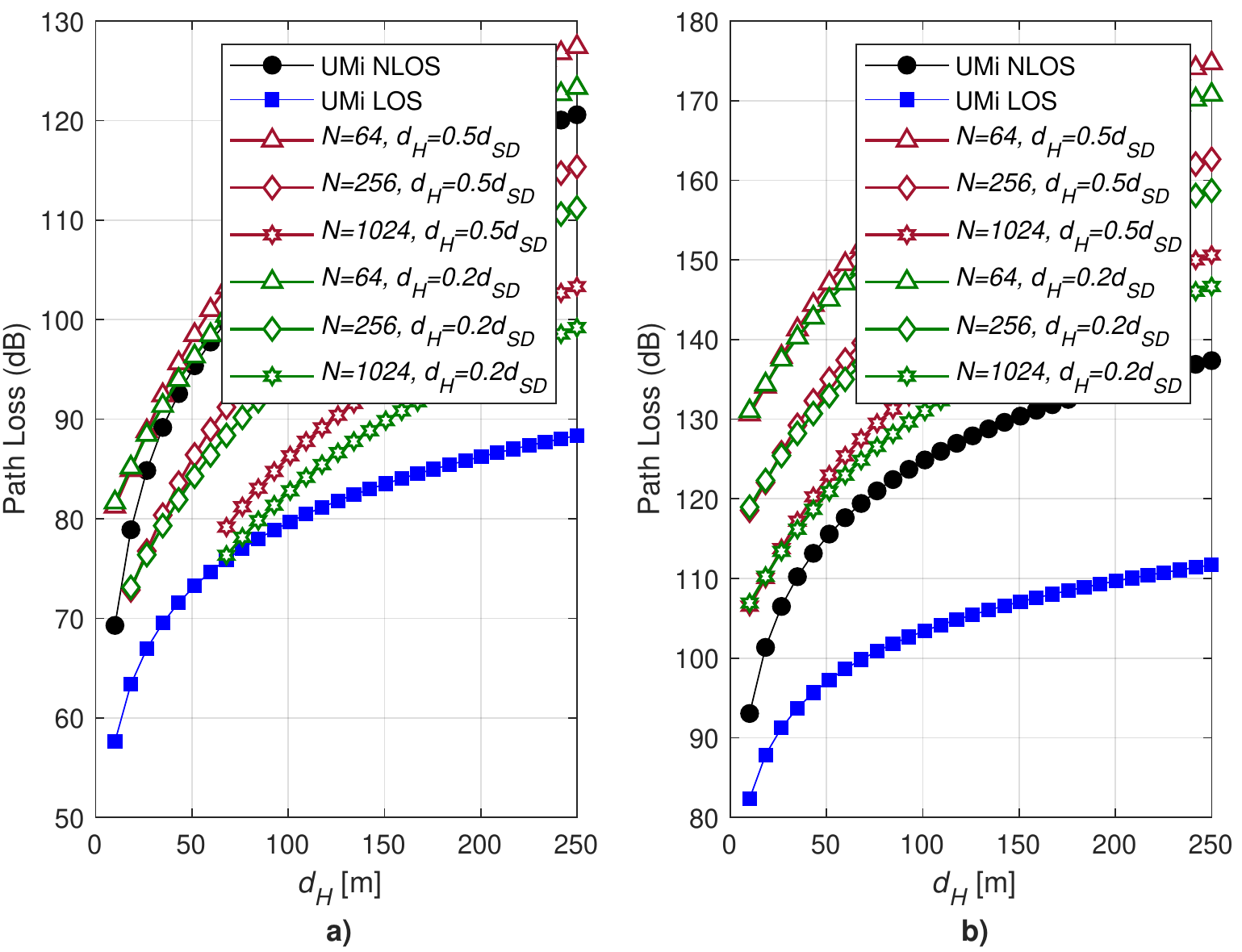}}
		\vspace*{-0.0cm}
		\caption{Path loss comparison of a single RIS path with varying $N$ and the direct transmission under (a) 3GPP UMi path loss model at $2.4$ GHz and (b) 5G UMi-Street Canyon path loss model at $28$ GHz.}\vspace*{-0.5cm}
		\label{fig:rissc}
	\end{center}
\end{figure}
	
For simplicity, we assume that RIS-elements reflect signal with unit-gain reflection coefficients \small($|\Gamma|=1$) \normalsize  and in a such way that all the signals coming through different RIS elements are aligned in phase at the receiver \small($\varphi_i=\phi_i$)\normalsize.  We also assume omni-directional, unit gain antennas in both transmitter receiver. Then \eqref{rec_pow_all_RIS} becomes:
	\begin{equation} \label{rec_pow_all_RIS2}
	p_{r_{i}} = \left(\sum\nolimits_{i} \sqrt{  \frac{p_t}{P_L(d_{SR})P_L(d_{RD})}}\right)^2
	\end{equation}
	
	Therefore the total path loss is given by:
	
	\begin{equation} \label{PL_all_RIS}
	P_{L_{TOT}} = \left(\sum\nolimits_{i} \sqrt{  \frac{1}{P_L(d_{SR})P_L(d_{RD})}}\right)^{-2}
	\end{equation}
	
	PLE ($ n $) is then calculated by employing log-distance path loss model \cite{Rappaport}:
	\begin{equation}\label{ple}
	P_L(d)\, [\text{dB}] = P_L(d_{0})+10n\log_{10}\left(\frac{d}{d_{0}}\right) 
	\end{equation}
	where $ d_0 $ is the reference distance, which is taken as $10$ m.

	The effect of an RIS on path loss is investigated for the scenario of Fig. \ref{single_RIS_model} for $d_V = 10$ m and $d_{SD}$ from $10$ m to $250$ m with $N = 64, 256$ and $1024$. We consider that the RIS is located both at the midway ($d_H = 0.5d_{SD}$) and closer to D ($d_H=0.2d_{SD}$). The power received through the NLOS channel is excluded in order to show the RIS effect clearly. The calculated total path loss of the RIS path is shown in Fig. \ref{fig:rissc} with LOS and NLOS UMi path loss models at $2.4$ and $28$ GHz. 
	
	We observe that path loss of the RIS-assisted scenario converges to the system modeled with UMi LOS when number of elements gets higher at lower frequencies. A comparable path loss to the NLOS system is obtained through the RIS path for number of elements greater than $ 64 $ at $2.4$ GHz, whereas more than $ 1024 $ elements are required for NLOS level path loss at $ 28 $ GHz. On the other hand, implementation of RISs with higher number of elements is cheaper at high frequencies due to smaller wavelengths. Locating RIS close to one of the terminals yields a minor improvement of $1$-$2$ dB on path loss depending on the distance between terminals over the midway location.  Calculated PLEs are listed in Table \ref{tab:PLtable}. Our results indicate improved PLEs with the RIS paths over the NLOS system model.
		
	\begin{figure}[t]
		\begin{center}\resizebox*{7.5cm}{6.4cm}{\includegraphics{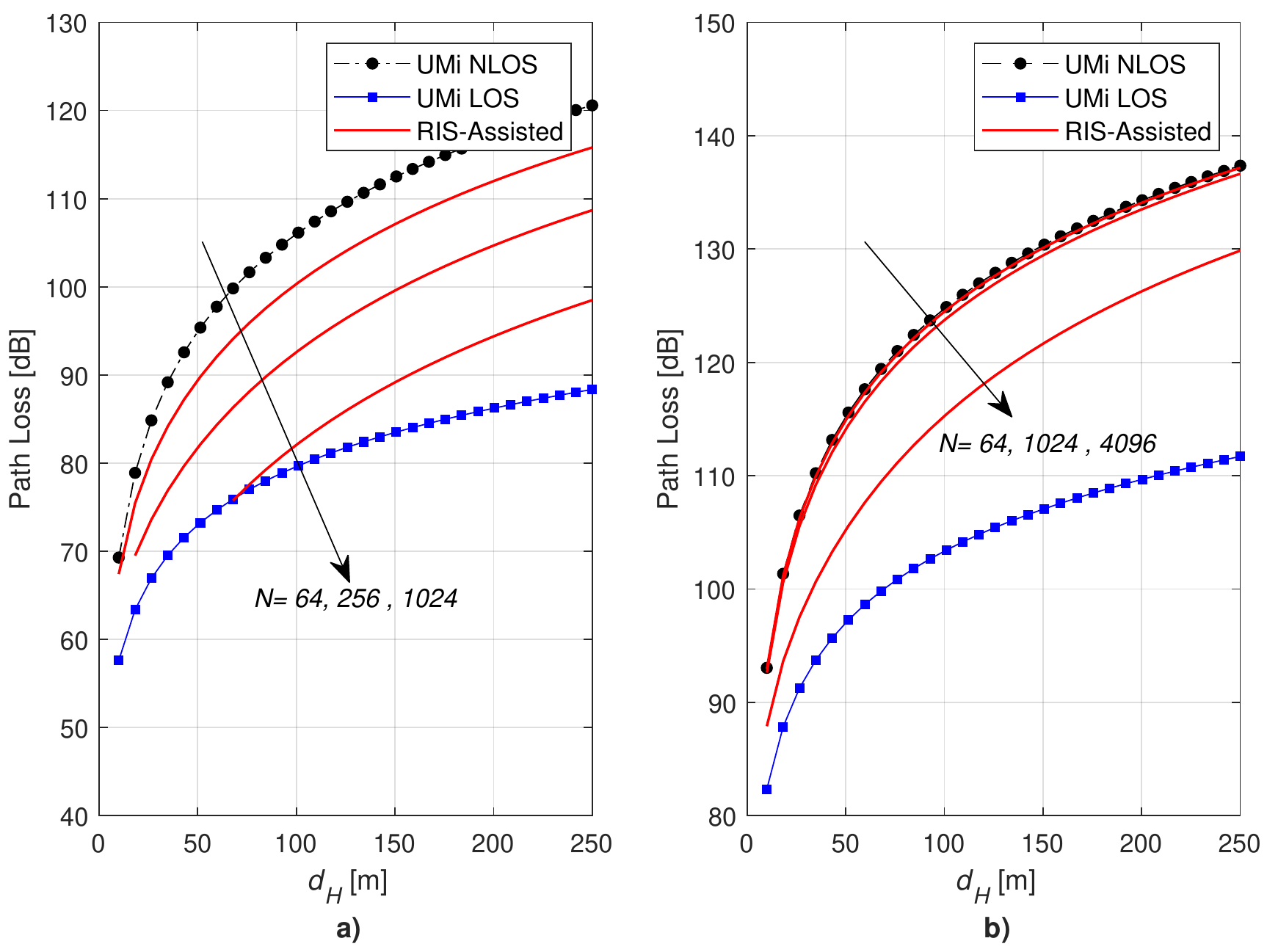}}
			\vspace*{-0.0cm}
			\caption{Path loss comparison of a single RIS-assisted system with varying $N$ and the direct transmission under (a) 3GPP UMi path loss model at $2.4$ GHz and (b) 5G UMi-Street Canyon path loss model at $28$ GHz.}\vspace*{-0.5cm}
			\label{fig:rissc2}
		\end{center}
	\end{figure}

\begin{table}[t] 
	\centering
	\caption{Path Loss Exponent of RIS Channels}
	\label{tab:PLtable}
	\begin{tabular}{lcccc} 
		\hline
		Frequency							& RIS Location		& PLE ($n$) \\ \hline
		\multirow{4}{*}{$2.4$ GHz} 				& No RIS, NLOS		  	& 3.67  \\ 
		& No RIS, LOS			& 2.2    \\ 	
		& $d_H = 0.5d_{SD}$   	& 3.08  \\ 
		& $d_H=0.2d_{SD}$		& 2.94    \\ \hline
		\multirow{4}{*}{$28$ GHz} 				& No RIS, NLOS		  	& 3.17  \\ 
		& No RIS, LOS			& 2.1    \\ 	
		& $d_H = 0.5d_{SD}$   	& 2.94  \\ 
		& $d_H=0.2d_{SD}$		& 2.80    \\ \hline
	\end{tabular} 
\end{table}

Total path loss of RIS assisted channel, including the RIS path and the NLOS path between the terminals from 3GPP UMi and 5G UMi-Street Canyon path loss models is shown in Fig. \ref{fig:rissc2} for RISs located at a distance of $0.2d_{SD}$. We observe greater reduction in the path loss at $2.4$ GHz in RIS-assisted channel over the NLOS system model when compared to $ 28 $ GHz communication channel. Calculated amount of reductions in the RIS-assisted path loss over the NLOS model ($\Delta P_L$) and the path loss exponents are listed in Table \ref{tab:PLtableAss}. $ 28 $ GHz channel reduces up to $3.5$ dB by use of $1024$ RIS elements, whereas $2.4$ GHz channel can be improved by an order of $23.8$ dB when $1024$ RIS elements are in use. RIS assisted channels demonstrate PLEs between NLOS and LOS system models for both frequencies, where as PLE reduces by increasing number of RIS elements. We observe that total path loss reduces inversely by the square of the number of RIS elements when RIS path gets dominant over the NLOS path, which is in agreement with (\ref{bep_bpsk1}) and \cite{Basar_2019_LIS} as long as the RIS path is dominant.

Our analysis shows that enhancement and control of path loss and PLE is achievable by use of an RIS in the presence of obstacles blocking the LOS path between the transmitter and the receiver. In other words, an RIS transforms the wireless propagation environment into a controllable entity even by reducing path loss in NLOS propagation environments. 

\begin{table}[t]
	\centering
	\caption{Path Loss Exponent of RIS Assisted Channels}
	\label{tab:PLtableAss} 
	\begin{tabular}{lcccc}
		\hline
		Frequency					&$N$	& $\Delta P_L$ (dB) & PLE ($n$) \\ \hline
		\multirow{3}{*}{$2.4$ GHz} 	&64    & 5.6				& 3.373  	\\ 
		&256	& 13.3				& 3.068   	\\  
		&1024 	& 23.8				& 2.847  	\\ \hline 
		\multirow{3}{*}{$28$ GHz} 	&64    & 0.3  				& 3.159 	\\ 
		&256	& 1 				& 3.130   	\\  
		&1024 	& 3.5				& 3.038  	\\ \hline                 
	\end{tabular}%
\end{table}

\begin{figure*}[t] 
	\begin{center}\resizebox*{17cm}{7cm}
		{\includegraphics[scale=1.1]{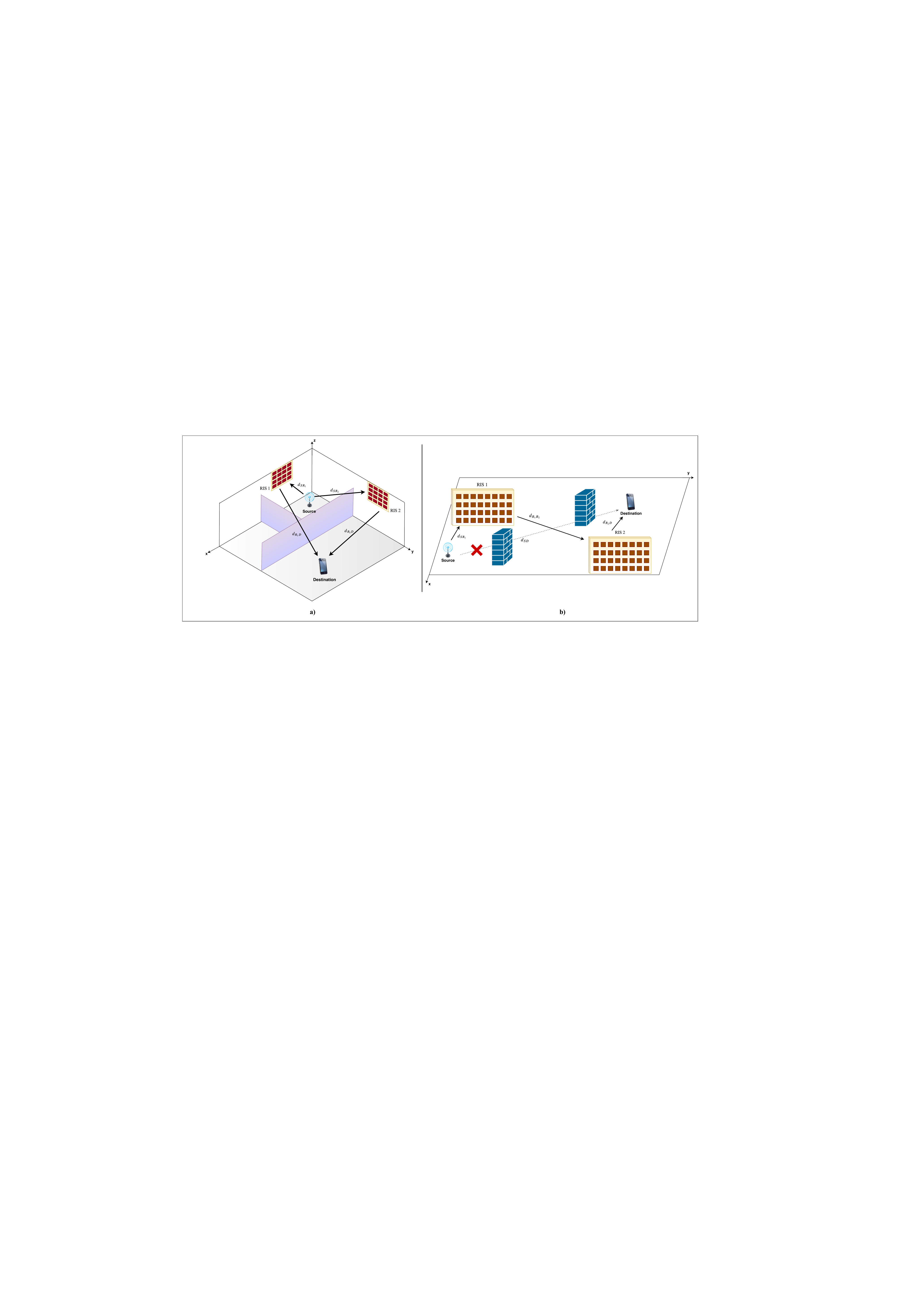}}
		\vspace*{-0.3cm}
		\caption{Communications over two RISs: (a) Indoor propagation scenario and (b) Outdoor propagation scenario (Each RIS is equipped with $N$ reflecting elements).} \label{fig:multiple}
		\vspace*{-0.3cm}
	\end{center}
\end{figure*}

\section{Wireless Communication Through Multiple RISs}

In future wireless sytems, the ubiquitous communications of many devices with various sizes are envisaged. However, it may not be reasonable to use multiple antennas or high-power consuming components in those devices to enable reliable and high-speed wireless transmission. Therefore, it becomes necessary to transfer this cost from wireless devices to the available RISs in propagation environments, where we can increase the number of RISs in a more flexible manner. It is worth noting that although many RIS-assisted systems have been investigated in recent times, the communication scenarios with multiple RISs have not been explored in a comprehensive manner so far. Within this perspective, the use of multiple RISs may have the potential to bring more promising advantages in terms of QoS, reliability and flexibility in the system design.

In this section, we conduct performance analysis of multiple RISs under two envisaged scenarios, namely, i) simultaneous transmission over two independent RISs in indoor environments and ii) double-RIS reflected transmission over a single link in outdoor environments. 
Under these multiple RIS-assisted transmission scenarios, we derive a generalized mathematical framework on the error performance which is valid for various path loss models.

\subsection{Case Study I: Indoor Communications with Multiple RISs}
In our first case study, we consider the three dimensional (3D) indoor communications model of Fig. \ref{fig:multiple}(a), where the direct path between S and D is blocked owing to the obstacles and the signal transmission is accomplished over two independent paths supported by two different RISs. This indoor scenario, which is given in 3D Cartesian coordinate system, includes typical open and closed large office environments with a maximum S-D separation of $100$ m.

In this setup, $d_{SR_k}$ and $d_{R_kD}$ respectively represent the distances of S-to-$k$th RIS and $k$th RIS-to-D for $k=1,2$. Furthermore, small-scale fading coefficients of the S-to-$k$th RIS and $k$th RIS-to-D channels are denoted by $h_i^{SR_k}$ and $h_i^{R_kD}$ for $i=1,2,...,N_k$ and $k=1,2$, respectively.  In this scenario,  RIS $1$ is located along the $x$-$z$ plane and has $N_1$ reflectors, while RIS $2$ is located along the $y$-$z$ plane with $N_2$ reflectors.

Under the condition of slow and flat Rician fading channels with Rician factor $K$, the baseband signal at D is given by 
 	\begin{align} \label{2indep_r}
	r=& \sqrt{p_t} \bigg[ \sqrt{P_{L}^{1}}\sum\nolimits_{i=1}^{N_1} {h_i^{SR_1}e^{j\phi_i^{(1)}}h_i^{R_1D} } \nonumber \\ +&\sqrt{P_{L}^{2}}\sum\nolimits_{j=1}^{N_2}{h_j^{SR_2}e^{j\phi_j^{(2)}}h_j^{R_2D} } \bigg] x+w
	\end{align}
where $P_L^{k}$ stands for path loss of $k$th RIS-assisted link and $\phi_i^{(k)}$ denotes the induced phase by $i$th reflecting element of $k$th RIS for $k=1,2$. The channels can be represented by $	h_i^{SR_1} = \alpha_{i}^{(1)}e^{-j\theta_{i}^{(1)}}$, $h_i^{R_1D} = \beta_{i}^{(1)}e^{-j\varphi_{i}^{(1)}}$, $h_j^{SR_2} = \alpha_{j}^{(2)}e^{-j\theta_{j}^{(2)}}$, and $h_j^{R_2D} = \beta_{j}^{(2)}e^{-j\varphi_{j}^{(2)}}$ in terms of their amplitudes and phases. Therefore, (\ref{2indep_r}) can be expressed as \small
\begin{align}
r=&\sqrt{p_t} \bigg[ \sqrt{P_{L}^{1}}\sum\nolimits_{i=1}^{N_1} \alpha_{i}^{(1)}\beta_{i}^{(1)}e^{j\Delta \Phi_1}  \nonumber \\ +&\sqrt{P_{L}^{2}}\sum\nolimits_{j=1}^{N_2}{\alpha_{j}^{(2)}\beta_{j}^{(2)}e^{j\Delta \Phi_2}}  \bigg] x+w 
\end{align} \normalsize
where $\Delta \Phi_1 =\phi_i^{(1)}-\theta_{i}^{(1)}-\varphi_{i}^{(1)} $ and $\Delta \Phi_2 =\phi_j^{(2)}-\theta_{j}^{(2)}-\varphi_{j}^{(2)} $ are phase difference terms.

The instantaneous SNR at D can be expressed as \small
\begin{align} \label{SNR_two_indep}
\gamma =\frac{{{\left|\sqrt{{P_{L}^{1}}} \sum\nolimits_{i=1}^{N} \alpha_{i}^{(1)}\beta_{i}^{(1)}e^{j\Delta \Phi_1}  +\sqrt{{P_{L}^{2}}}\sum\nolimits_{j=1}^{N}{\alpha_{j}^{(2)}\beta_{j}^{(2)}e^{j\Delta \Phi_2}} \right|}^{2}}{p_t}}{N_0}.
\end{align} \normalsize
In (\ref{SNR_two_indep}), the SNR is maximized with the phase alignment satisfying $\Delta \Phi_1=\Delta \Phi_2=0$, for all $i=1,2,...,N_1$ and $j=1,2,...,N_2$; therefore, the maximized SNR can be obtained as in (4): \small
\begin{align} \label{SNR_max_indoor}
\gamma_{\max} =\frac{{{\left|\sqrt{P_{L}^{1}} \sum\nolimits_{i=1}^{N_1} \alpha_{i}^{(1)}\beta_{i}^{(1)} +\sqrt{P_{L}^{2}}\sum\nolimits_{j=1}^{N_2}{\alpha_{j}^{(2)}\beta_{j}^{(2)}} \right|}^{2}}{p_t}}{N_0} =\frac{A^2{p_t}}{N_0}.
\end{align} \normalsize
It should be noted that $\alpha_i^{(k)}$ and $\beta_i^{(k)}$ ($k=1,2$) are independent and follow Rician distribution. In \eqref{SNR_max_indoor}, the random variable $A$ is the sum of $N_1$ and $N_2$ random variables. If $N_1\gg 1$ and $N_2 \gg 1$, $A$ follows Gaussian distribution, and its mean and variance are respectively given as follows:  
\begin{align} \small
&\text{E}[A]=\frac{\left(N_1\sqrt{P_L^1}+N_2\sqrt{P_L^2}\right)\pi }{4(K+1)} L_{1/2}^2\left(-\frac{K^2}{(K+1)} \right) , \nonumber\\
&\text{VAR}[A] =\left( N_1P_L^1+N_2P_L^2\right) \bigg( 1-\frac{\pi^2 L_{1/2}^4\left(-\frac{K^2}{(K+1)} \right) }{16(K+1)^2}\bigg).
\end{align}\normalsize
 It is worth noting that $\gamma$ will have the same distribution regardless of the type of small-scale fading due to the CLT.
In order to show effectiveness of our analyses, using the MGF approach and following the same steps as in Section II with (8), upper bounded SEP for BPSK can be obtained as
\begin{align} \label{bpsk_upper_indoor}
P_e \leq    \frac{0.5\exp\left( \frac{-\frac{\left(N_1\sqrt{P_L^1}+N_2\sqrt{P_L^2} \right)^2\pi^2L_{1/2}^4\left(-K^2/(K+1)\right) \it{p_t}}{16(K+1)^2N_0}}{1+\frac{\left(N_1P_L^1+N_2P_L^2\right)\left[ 16(K+1)^2-\pi^2L_{1/2}^4\left(-K^2/(K+1)\right)\right]\it{p_t}}{8(K+1)^2N_0}} \right)  }{\sqrt{1+\frac{\left(N_1P_L^1+N_2P_L^2\right)\left[ 16(K+1)^2-\pi^2L_{1/2}^4\left(-K^2/(K+1)\right)\right]\it{p_t}}{8(K+1)^2N_0}}} . 
\end{align}

We can generalize this simultaneous transmission scenario with two RISs to the general case of $ N_S $ independent RISs. If each independent RIS consist of $ N $ reflector elements and path losses of all paths are approximately equal with $ {P_{L}^{i}}={P_{L}}$ ($i=1,2, \dots, N_S$), the generalized SEP is expressed as follows for BPSK: 
\begin{align} \label{bpsk_upper_indoor}
P_e \leq    \frac{0.5\exp\left( \frac{-\frac{N_S^2N^2P_L \pi^2L_{1/2}^4\left(-K^2/(K+1)\right) \it{p_t}}{16(K+1)^2N_0}}{1+\frac{N_SNP_L\left[ 16(K+1)^2-\pi^2L_{1/2}^4\left(-K^2/(K+1)\right)\right]\it{p_t}}{8(K+1)^2N_0}} \right)  }{\sqrt{1+\frac{N_SNP_L\left[ 16(K+1)^2-\pi^2L_{1/2}^4\left(-K^2/(K+1)\right)\right]\it{p_t}}{8(K+1)^2N_0}}} . 
\end{align}
While the error performance is inversely proportional to \small$\left(N_1\sqrt{P_L^1}+N_2\sqrt{P_L^2} \right)^2$\normalsize in the two RIS-assisted transmission, the simultaneous transmission in the presence of $ N_S $ different RISs provides 
\begin{align} \small \label{bep_simul_2}
P_e \propto {\exp{\left( -\frac{N_S^2N^2P_L \pi^2L_{1/2}^4\left(-K^2/(K+1)\right) \it{p_t}}{16(K+1)^2N_0} \right) }} 
\end{align}\normalsize
where $N_SN$ is the total number of reflectors in the system. In (\ref{bep_simul_2}), it is observed that the error performance is inversely proportional to $(N_SN)^2$, which provides more flexibility in design when the RIS sizes are limited in indoor environments. In other terms, instead of employing a single large RIS, the same performance can be provided by multiple smaller RISs, which is better suited to indoor applications.  
\subsection{Case Study II: Outdoor Communications with Multiple RISs}

The system model of the double-RIS reflected transmission in outdoor propagation environment is illustrated in Fig. \ref{fig:multiple}(b). Here, since the multi-hop RISs-assisted scenarios will deteriorate the received signal power due to the multiplicative path loss, the available RISs have been placed in parallel instead of serial to prevent the loss in the received signal. The outdoor scenario, which is modeled in 2D Cartesian coordinates, includes typical dense urban environments with a minimum S-D separation above $100$ m. In this scenario, since the direct path between S and D is blocked, the signal is transmitted via two RISs located near to S and D. This system can be used to alleviate the shortcomings of the use-cases that may arise particularly in outdoor environments in $6$G and beyond communications systems, especially in mmWave and THz bands. For instance, if it is desired to convey data in an outdoor environment over a long-distance, at high frequencies, the signal can be attenuated or lost due to many obstacles in between. In this case, the nearest RISs, which are equipped with $ N $ reflecting elements, are chosen by S and D for transmission, and communication is carried out by these RISs. We later show that using two RISs, the wireless environment is transformed into a virtual MIMO system. 

The double-RIS reflected transmission scenario includes a single path from S to D. The channels between S-to-RIS 1 and RIS 2-to-D can be modeled as deterministic LOS channels while the channel between $i$th element of RIS 1-to-$j$th element of RIS 2 $h^{R_1R_2}_{ij}$ is modeled by Rician fading with $K$ factor for $i=1,2,...,N$ and $j=1,2,...,N$. Distance of the S-to-RIS 1, RIS 1-to-RIS 2 and RIS 2-to-D is respectively denoted by $d_{SR_1}$, $d_{R_1R_2}$ and $d_{R_2D}$. $d_{SR_1}$ and $d_{R_2D}$ are sufficiently small such that small-scale fading can be ignored. From the standpoint of channel amplitudes and phases, $h^{R_1R_2}_{ij}$ can be expressed as $h^{R_1R_2}_{ij}=\beta_{ij}e^{-j\varphi_{ij}} $.   

\begin{figure*}[t] 
	\begin{center}\resizebox*{17cm}{7cm}
		{\includegraphics[scale=1.1]{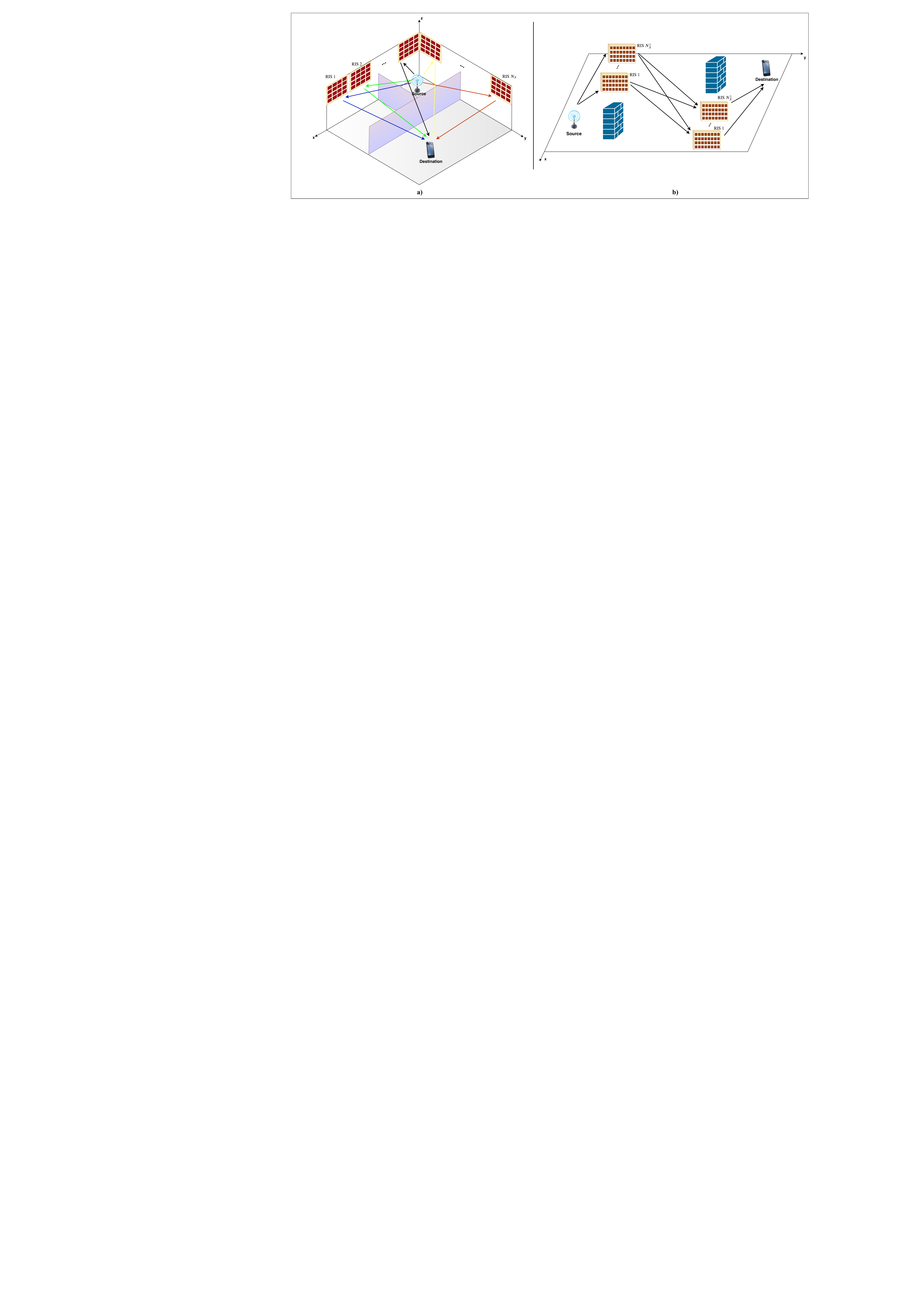}}
		\vspace*{-0.3cm}
		\caption{RIS-selection based transmission over multiple RISs: (a) Indoor propagation scenario and (b) Outdoor propagation scenario .} \label{fig:multiple_genel}
		\vspace*{-0.3cm}
	\end{center}
\end{figure*}

The baseband signal at D is obtained as follows: \small
\begin{align} \label{eq:outdoor_r}
r= \sqrt{{p_t}P_{L}^R}\left(\sum\nolimits_{i=1}^{N} \sum\nolimits_{j=1}^{N}{e^{j\phi_i^{(1)}}h^{R_1R_2}_{ij} e^{j\phi_j^{(2)}}} \right)x+w \vspace*{-0.3cm}
\end{align} \normalsize
where $\phi_i^{(k)}$ is as defined in \eqref{2indep_r}. Then, the instantaneous SNR at D can be easily calculated as
\begin{align}
\gamma =\frac{{{\left| \sqrt{P_{L}^R}  \sum\nolimits_{i=1}^{N} \sum\nolimits_{j=1}^{N}{ \beta_{ij} e^{j(\phi_i^{(1)}+\phi_j^{(2)}-\varphi_{ij})} } \right|}^{2}}{p_t}}{N_0}.
\end{align}
SNR at D is maximized with the phase elimination
($\phi_i^{(1)}+\phi_j^{(2)}=\varphi_{ij}$ for $i=1,2,...,N$ and $j=1,2,...,N$). Consequently, the maximized SNR is obtained as \small
\begin{align} \label{eq:max_outdoor}
\gamma_{\max} =  \frac{{{\left| \sqrt{P_{L}^R}  \sum\nolimits_{i=1}^{N} \sum\nolimits_{j=1}^{N}{ \beta_{ij} } \right|}^{2}}{{p_t}}}{{{N}_{0}}}
=\frac{A^2{p_t}}{N_0}
\end{align} \normalsize
where $A$ follows Gaussian distribution with \small$N^2\sqrt{P_L^R}\sqrt{\pi/(4(K+1))}L_{1/2}\left(-K^2/(K+1)\right)$ \normalsize  mean and \small$N^2 P_L^R \left(1-\pi/(4(K+1))L_{1/2}^2\left(-K^2/(K+1)\right)\right)$ \normalsize variance due to the CLT for $N \gg 1$. Then, following the same steps as in previous sections, the upper bounded SEP for BPSK can be obtained using (8) as
\begin{align} \label{bep_aligned}
P_e \leq \frac{1}{2} \left(  \frac{\exp{\left(- \frac{\frac{N^4P_L^R\pi L_{1/2}^2\left(-K^2/(K+1)\right) p_t}{4(K+1)N_0}}{1+\frac{N^2P_L^R\left[4(K+1)-\pi L_{1/2}^2\left(-K^2/(K+1)\right)\right] p_t}{2(K+1)N_0}}\right) }}{\left( 1+\frac{N^2P_L^R\left[4(K+1)-\pi L_{1/2}^2\left(-K^2/(K+1)\right)\right]p_t}{2(K+1)N_0}\right)^{1/2}} \right).
\end{align}
As can be seen from (\ref{bep_aligned}), in the low SNR region, the error performance of the double-RIS reflected transmission systems ($P_e$) can be approximated by 
\begin{align} \label{bep_aligned_2}
P_e \propto {\exp{\left( -\frac{N^4P_L^R\pi L_{1/2}^2\left(-K^2/(K+1)\right) p_t}{4(K+1)N_0} \right) }}. 
\end{align}
The term $ N ^ 4 $ in (\ref{bep_aligned_2}) brings a significant improvement in error performance due to the virtual $N \times N$ MIMO channel created between RIS 1 and RIS 2. It should be also noted that \eqref{bep_aligned} can be used with various path loss models directly, while for other small-scale fading models, only slight modifications are required using the MGF approach due to the CLT. While the employing two serial RISs deteriorates the received signal power, it will be possible to maintain a reliable transmission particularly in systems with low power consumption and for cell-edge users by increasing the number of RIS elements and positioning the available RISs appropriately.
%

\subsection{RIS Selection Strategies}
 Although it is possible to improve the QoS using multiple RISs in future wireless networks, this will increase the overall complexity and render simultaneous phase adjustment a challenging task. By utilizing RIS selection over multiple RISs, a low-complexity and cost-effective transmission can be provided while many advantages of multiple RIS-assisted systems are preserved. For this purpose, we propose the novel concept of RIS selection for the communication scenarios of Fig. \ref{fig:multiple_genel}.  

As shown in Fig. \ref{fig:multiple_genel}(a), $ N_S $ RISs are placed in an indoor propagation environment, where each RIS has $N$ reflecting elements. Here, $d_{SR_k}$ and $d_{R_kD}$ respectively represent the distances of S-to-$k$th RIS and $k$th RIS-to-D for $k=1,2, \dots,K$. Furthermore, small-scale fading coefficients of the S-to-$k$th RIS and $k$th RIS-to-D channels are denoted as in (17) by $h_i^{SR_k}=\alpha^{(k)}_ie^{-j\theta^{(k)}_i}$ and $h_i^{R_kD}=\beta^{(k)}_ie^{-j\varphi^{(k)}_i}$ for $i=1,2,\dots,N$ and $k=1,\dots,N_S$, respectively. The received SNR of the signal transmitted through only the $k$th RIS ($\gamma_k$) is calculated as
\begin{align} \label{SNR_indoor}
\gamma_k \!=\!\frac{{{\left|\!  \sum\nolimits_{i=1}^{N}{\sqrt{P_{L}^i}\alpha^{(k)}_i\beta^{(k)}_ie^{j\Delta \Phi^k_i}}  \right|}^2}{\it{p_t}}}{N_0}.
\end{align} 
where $\Delta \Phi_k =\phi_i^{(k)}-\theta_{i}^{(k)}-\varphi_{i}^{(k)} $ is the phase difference for the $k$th path. Considering the maximized SNR values as in (18), an RIS selection is conducted over $N_S$ RISs to choose the RIS, which has highest SNR, for transmission as
 \begin{align} \label{SNR_indoor_max}
 \gamma'_{\text{In}} \!=\text{max}(\gamma_{\max,1},\gamma_{\max,2},\dots,\gamma_{\max,k})
 \end{align} 
 where $\gamma'_{\text{In}}$ is the maximized received SNR for RIS selection based system in indoor environment and $\gamma_{\max,k}$ is the maximized SNR of the $k$th path with proper phase adjustment. Here, we focus on the selection of a single RIS while a generalization might be straightforward. 
 
The generalized system model of the RIS selection based outdoor communication system is illustrated In Fig. \ref{fig:multiple_genel}(b). In this model, we assume $ N_S^1 $ and $ N_S^2 $ RISs, respectively, within the proximity of the transmitter and receiver. Here, the distances of S-to-RIS $k$, RIS $k$-to-RIS $l$ and RIS $l$-to-D are respectively denoted by $d_{SR_k}$, $d_{R_kR'_l}$ and $d_{R'_lD}$ for $k=1,\dots,N_S^1$ and $l=1,\dots,N_S^2$ as in \eqref{eq:outdoor_r}. In this setup, $N_S^1\times N_S^2$ possible transmission paths are available for communications. By selecting the path that has the highest SNR, the overall cost can be reduced using a single RIS for each terminal. As in \eqref{eq:max_outdoor}, the maximized SNR is obtained for each path as $\gamma^{k,l}_{\max}$. Therefore, RISs selection for outdoor environment is conducted over $N_S^1\times N_S^2$ paths as
 \begin{align} \label{SNR_outdoor_max}
\gamma'_{\text{Out}} \!=\underset{k,l}{\max}\ (\gamma^{k,l}_{\max})
\end{align} 
where $\gamma'_{\text{Out}}$ is the maximized received SNR for RIS selection based systems in outdoor environment.
 
\section{Numerical Results}

In this section, we used simulation models and studied the effects of various RIS applications taking into account path losses and channel fadings.   For all simulations in this section, path loss is modeled as in \eqref{ellingston_PL}, such as \small${{P}^{R}_{L}} \propto (d_{SR}d_{RD})^{-2}$\normalsize. The common parameters for simulations are set up as follows: $K=10$ dB, $G_T=1$, $G_R=1$, $G_e^i=G_e^r\cong5$ dBi as in \cite{Ellingson}.

\begin{figure}[t] 
	\begin{center}\resizebox*{7.5cm}{6cm}{\includegraphics{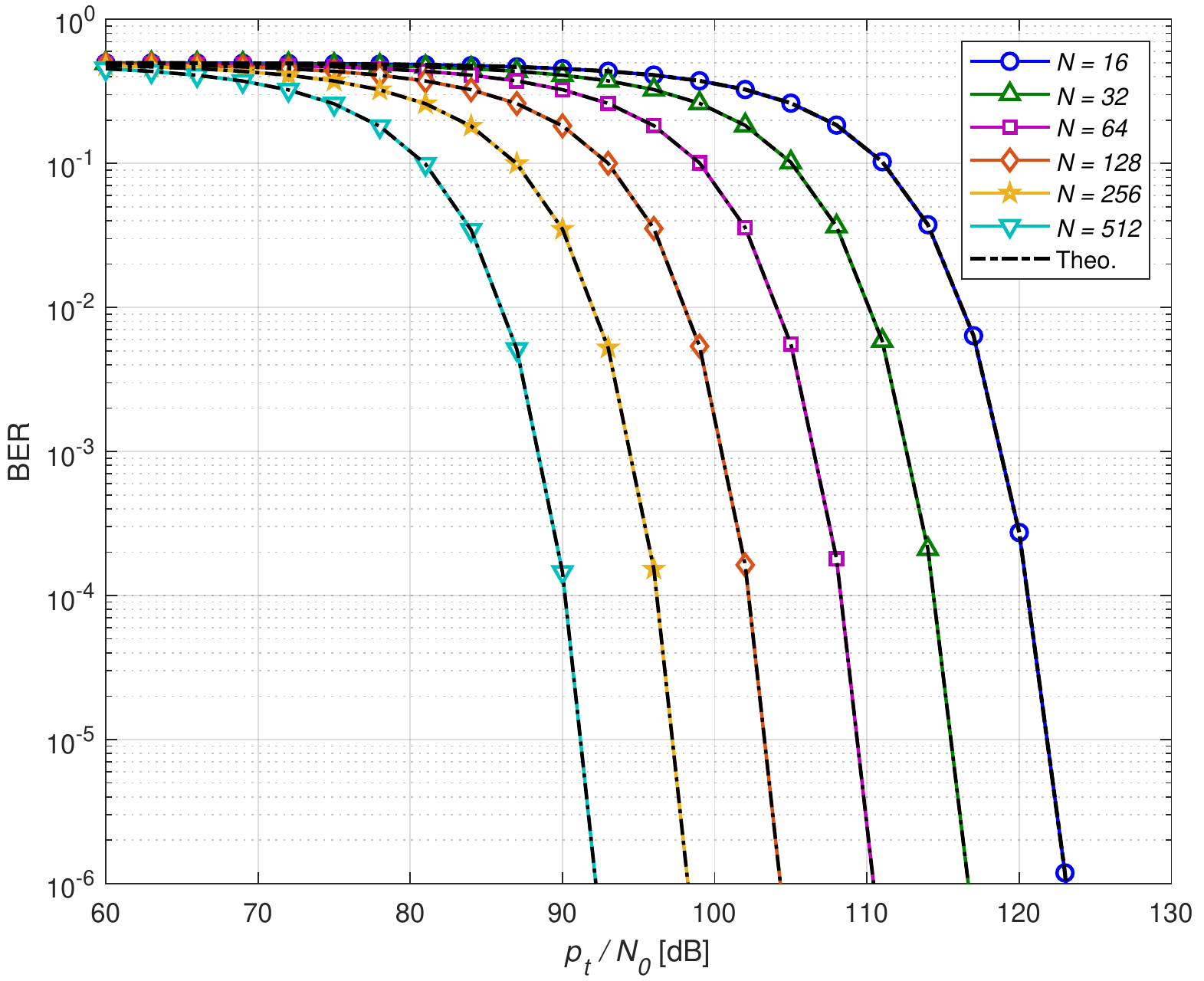}}
		\vspace*{-0.0cm}\caption{BER performance of the single-RIS assisted system for BPSK signaling and varying $N$ values.}\vspace*{-0.5cm} \label{fig:an_RIS_BER}
	\end{center}
\end{figure}

In Fig. \ref{fig:an_RIS_BER}, we show the bit error rate (BER) of the single-RIS assisted scheme in Fig. 1 versus $p_t/N_0$, where $p_t$ is the power consumed for transmission. Here, the simulations are conducted under $f_c=2.4$ GHz, $d_{SR}=25$ m and $d_{RD}=75$ m. In order to make a fair comparison, we assume that the equal power is consumed for all systems ($p_t=1$) and BPSK is employed. As clearly seen from Fig. 7, the simulation and exact analytical BER results in \eqref{bep_bpsk1} are in close agreement for high $N$ values due to the CLT. By doubling the number of reflectors, approximately a $6$ dB improvement in the required SNR is achieved, $P_e$ is inversely proportional to $ N ^ 2 $ as observed in (\ref{bep_bpsk1}).

\begin{figure}[t]
	\begin{center}\resizebox*{7.5cm}{6cm}{\includegraphics{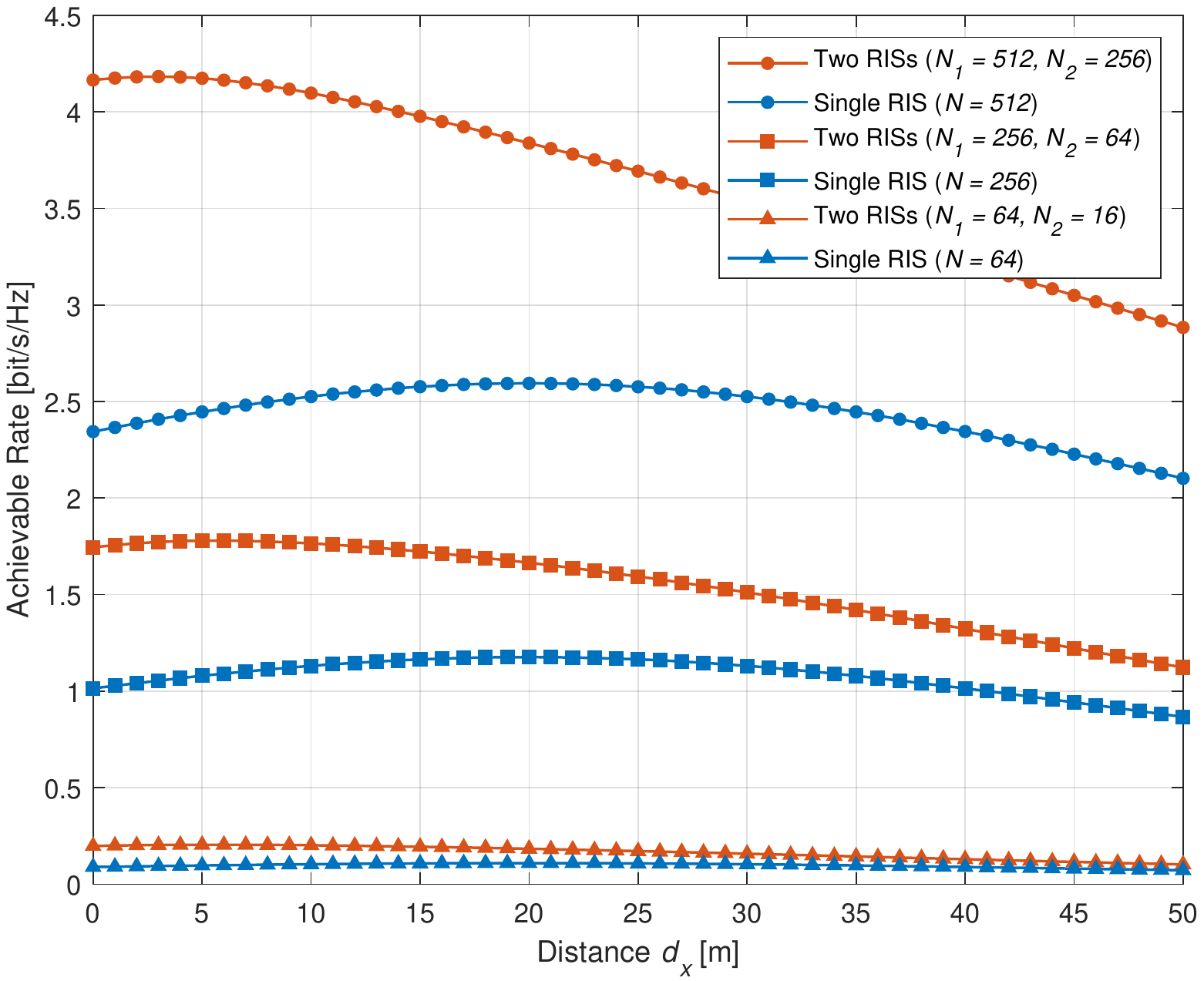}}
		\vspace*{-0.0cm}\caption{Comparison of achievable data rate performance for the two simultaneous RISs assisted system with a single-RIS assisted system in an indoor transmission scenario.}\vspace*{-0.5cm} \label{fig:rate_Two_single}
	\end{center}
	
\end{figure}

In Fig. \ref{fig:rate_Two_single}, the effect of using two simultaneous RISs on the achievable data rate in an indoor office environment with the dimensions of $50$ m$\times$$50$ m$\times$$10$ m  is investigated. In this comparison, achievable data rate analysis of two RISs-assisted and single RIS-assisted systems  are performed. Considering the transmission scenario in Fig. \ref{fig:multiple}(a), the first RIS with $N_1$ reflectors is located at $(20,0,10)$, while the second RIS with $N_2$ reflectors is located at $(0,25,10)$. Assuming that the source, which is located at $(5,5,0)$, transmits information to an user by using RISs-assisted LOS path.  The position of the user changes along the $x$-axis as $(d_x,40,0)$. For the single-RISs-assisted scenario, only the first RIS is activated during transmission and results are obtained for both systems at $30$ GHz operating frequency. Thus, a significant increase in achievable data rate is obtained for changing locations of the user by activating two simultaneous RISs during transmission.

\begin{figure}[t]
	\begin{center}\resizebox*{7.7cm}{6.3cm}{\includegraphics{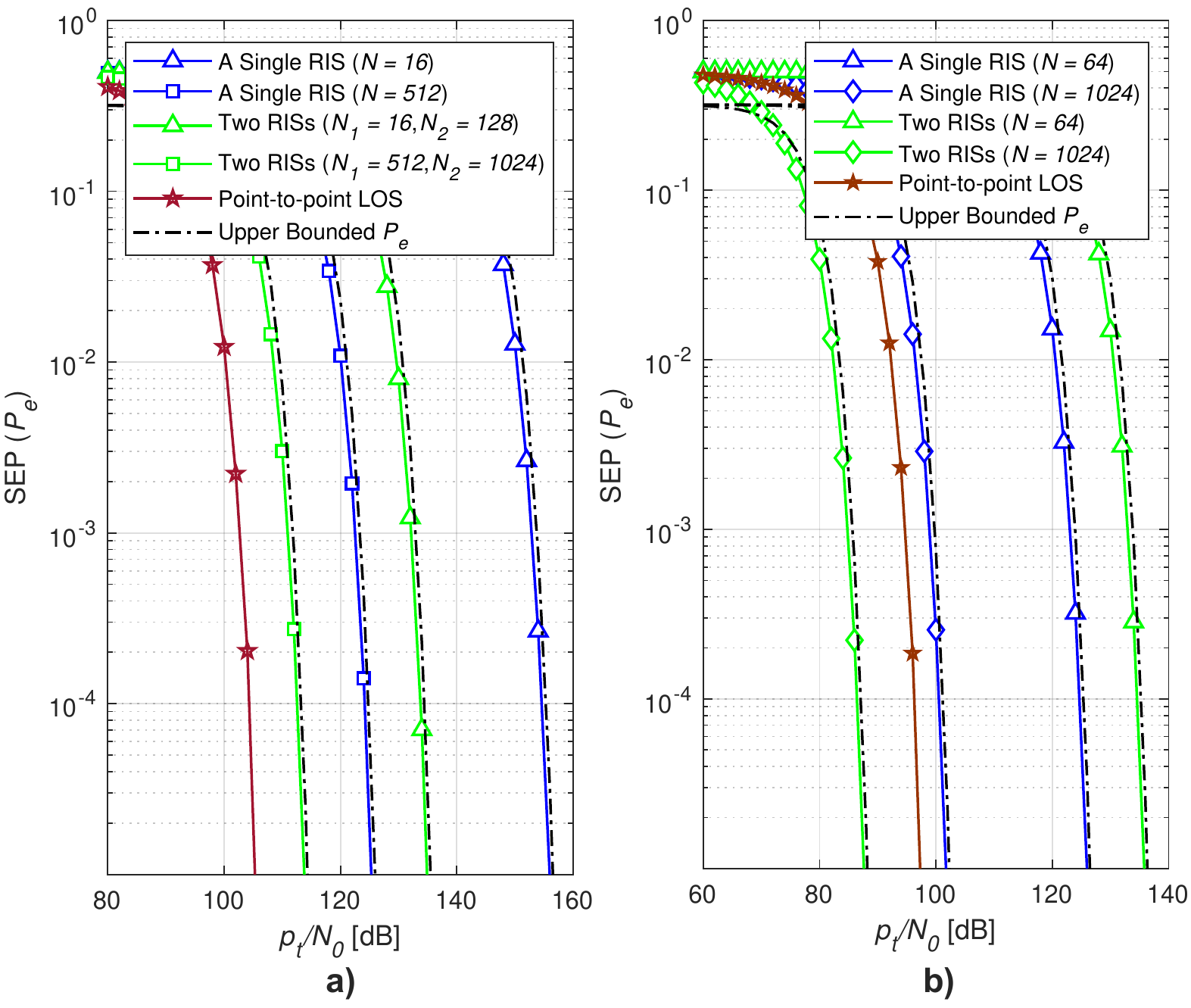}}
		\vspace*{-0.0cm}\caption{Comparison of exact and upper bounded average SEP for the RIS-assisted scenarios in (a) indoor environment and (b) outdoor environment under BPSK.}\vspace*{-0.5cm} \label{fig:BER_multiple}
	\end{center}
\end{figure}
In Fig. \ref{fig:BER_multiple}(a), the analytical error performance of systems with single and two RISs in an indoor environment is examined by making comparisons with a reference system utilizing only the direct point-to-point LOS transmission without small-scale fading. The distances are taken as $d_{SD}=50$ m in the point-to-point LOS system, $d_{SR}=20$ m, $d_{RD}=40$ m  in the single RIS-assisted system, and $d_{SR_1}=10$ m, $d_{SR_2}=15$ m, $d_{R_1D}=40$ m, $d_{R_2D}=35$ m in two simultaneous RIS-assisted setup.
$ 30 $ GHz operating frequency is considered for this comparison. In addition to increasing the data rate, when two RISs are activated simultaneously, it is possible to achieve significant performance increases in the SEP. When $ N_1 $ and $ N_2 $ are increased sufficiently, it is theoretically possible to achieve the performance of point-to-point LOS communication. In Fig. \ref{fig:BER_multiple}(b), we demonstrate the transformative effect of using multiple RISs on the propagation medium. In this comparison, SEP analysis of the double RIS reflected system as illustrated in Fig. \ref{fig:multiple}(b) and a single RIS-assisted system in an outdoor environment at $2.4$ GHz operating frequency. Here, the point-to-point LOS transmission scenario without fading is considered as a reference. The distances are taken as $d_{SD}=245$ m in the point-to-point LOS system, $d_{SR}=75$ m, $d_{RD}=165$ m  in the single RIS-assisted system, and $d_{SR_1}=20$ m, $d_{R_1R_2}=200$ m, $d_{R_2D}=20$ m in double RIS reflected transmission setup.
As clearly seen, even a single RIS with sufficiently large number of reflecting elements RIS can provide an effective solution
even if the transmission link is blocked. Furthermore, the double-RIS system provides a better error performance than the other setups when $N=1024$, since $P_e$ is inversely proportional to the fourth power of the number of reflectors \small($P_e \propto N^{-4}$) \normalsize while we have \small$P_e \propto(2N)^{-2}$ \normalsize in simultaneous transmission scenario over two RISs and \small$P_e \propto N^{-2}$ \normalsize in a single RIS-assisted system for low $p_t/N_0$ values. It should be noted that the double-RIS reflected scheme is the most challenging scheme in its design while offering a far better error performance than the others.  In other words, it is possible to convert a fading channel to a superior communication channel that acts as a non-fading one by intelligent reflections. It can be also stated that the direct path is transformed into an $N \times N$ virtual MIMO channel with the help of RISs, located close to the transmitter and receiver and each of which contains $ N $ reflectors in the double-RIS reflected transmission setup.

\begin{figure}[t]
	\begin{center}\resizebox*{7.7cm}{6.3cm}{\includegraphics{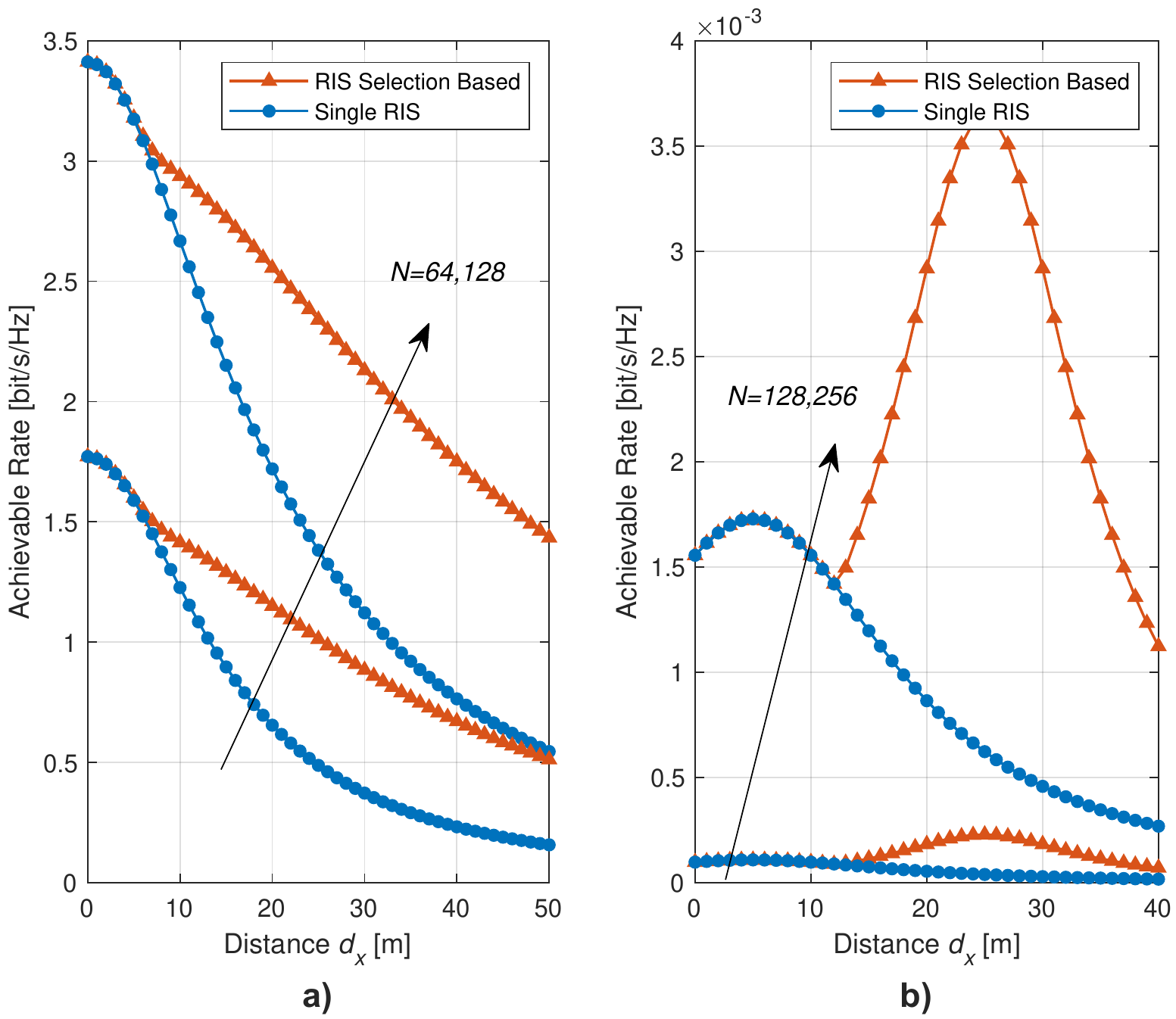}}
		\vspace*{-0.0cm}\caption{Comparison of achievable data rate performance for RIS selection based systems with single and multiple RIS-assisted systems in (a) indoor environment  and (b) outdoor environment.}\vspace*{-0.5cm} \label{fig:RIS_selection}
	\end{center}
\end{figure}
In Fig. \ref{fig:RIS_selection}, we implement RIS selection strategies for indoor and outdoor scenarios depicted in Fig. \ref{fig:multiple_genel}. First, we consider the indoor transmission with three available RISs by selecting the best one from \eqref{SNR_indoor_max}. Instead of a single RIS-assisted transmission, a best path can be selected among the three possible RIS-assisted paths with the following coordinates of RISs: \small$(0,35,10)$, $(20,0,10)$, $(0,15,10)$\normalsize.
It is assumed that all RISs are equipped with $N$ reflectors due to the simplicity and $p_t = 5$ W for all systems. The source, which operates at $28$ GHz, and user is respectively located at \small$(5,5,0)$ \normalsize and \small$(d_x,40,0)$\normalsize. As shown in Fig. \ref{fig:RIS_selection}(a), the RIS selection based transmission provides a better data rate performance than single RIS-assisted systems when the direct path is blocked for $N=64$ and $128$. Furthermore, the RIS selection strategy enables a flexible transmission in outdoor propagation environment when the direct path is blocked. When two RISs are available for both the transmitter and the receiver \small($N_S^1=2$ and $N_S^2=2$) \normalsize as illustrated in Fig. \ref{fig:multiple}(b), transmission can be carried out over four possible links \small($N_S^1\times N_S^2=4$)\normalsize. In this case, the RIS pairs that provide the highest SNR value are selected by using \eqref{SNR_outdoor_max} and transmission is provided via a single path. Using the system models in \eqref{eq:outdoor_r}, the coordinates along the $x$-$y$ plane for the four RISs are assumed as follows: $(0,15)$, $(20,10)$, $(5,215)$, $(25,220)$. The source and user are respectively located at $(0,0)$ and $(d_x,230)$. The results are obtained by assuming that the user moves along the $ x $-axis. The system which always use a same path for transmission is considered as a reference system for the comparison. As shown in Fig. \ref{fig:RIS_selection}(b), an improved achievable data rate performance is obtained under RISs selection with respect to double reflected two-RISs assisted systems without selection while the overall system complexity and phase adjustment costs are decreased.  RIS selection based system offers a flexible mechanism by choosing the most reliable path under varying locations and distances.  Here, under the changing position of the user on the $ x $-axis, the best performance is expected to be observed in the regions where the user is close to one of the RISs. Therefore, the highest achievable rate performance is seen where the user is located at $ 5 $ and $ 25 $ m on the $ x $-axis. Moreover, the most dominant factor in the RIS selection is the path loss rather than small-scale fading. It should be noted that the best way to ensure a reliable transmission is choosing the shortest path for transmission under same fading conditions.

\begin{figure}[t]
	\begin{center}\resizebox*{7.7cm}{6.3cm}{\includegraphics{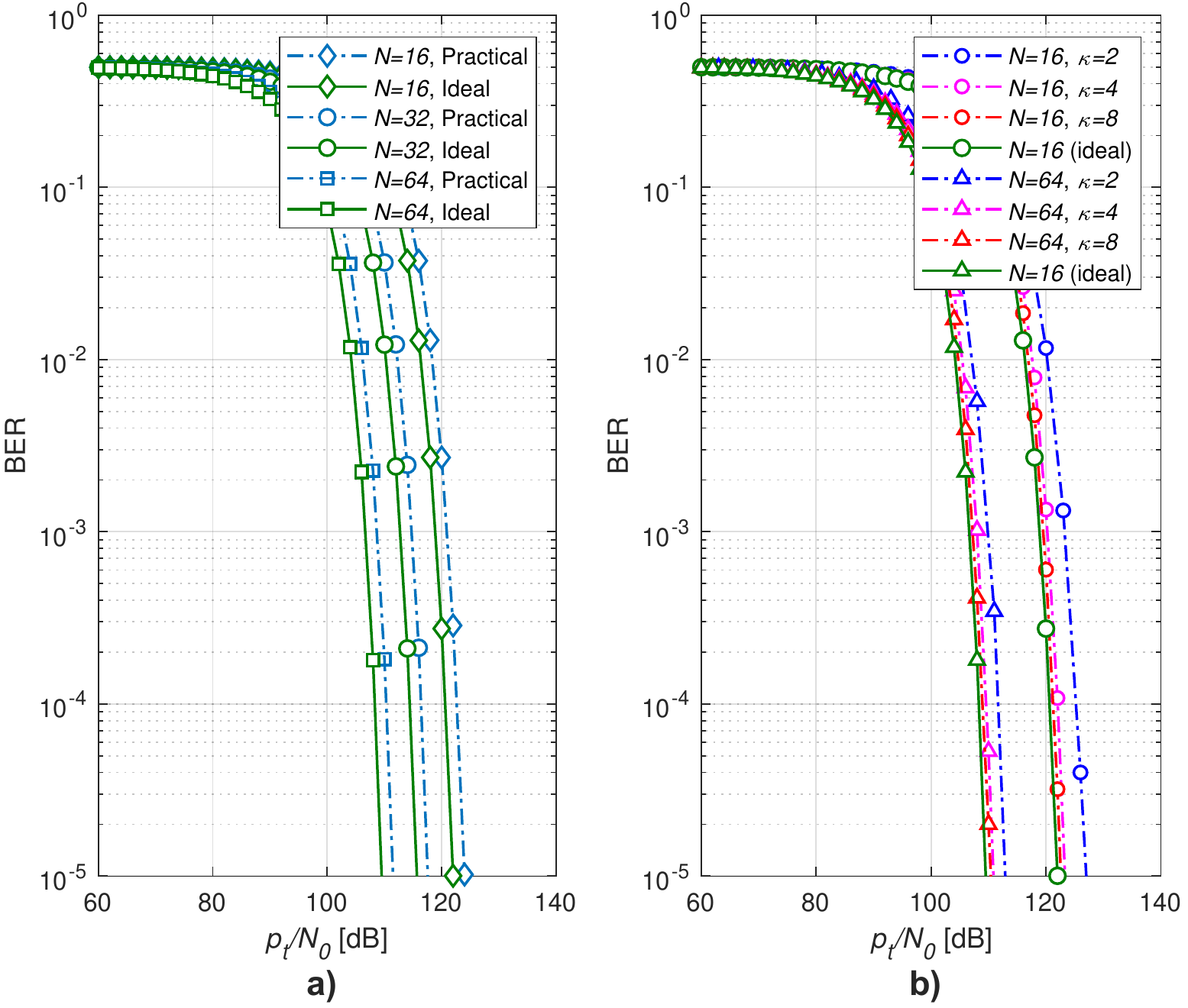}}
		\vspace*{-0.0cm}\caption{BER performance evaluations under the non-ideal conditions: (a) Transmission with practical reconfigurable meta-surface realization and (b) Transmission under imperfect channel phase knowledge.}\vspace*{-0.5cm} \label{non_ideal}
	\end{center}
\end{figure}

In order to give useful insights into the practical implementation of RISs, we consider  the practical reconfigurable metasurface designed in \cite{MDR-16}.  This metasurface is capable of adjusting reflection phases from $-150^{\circ}$ to $140^{\circ}$ for $|\Gamma|=-1$ dB. In Fig. \ref{non_ideal}(a), under a single RIS-assisted scenario, an error performance comparison is given for this RIS and the ideal one, which is capable of reflecting the signal within the $ [-180^\circ, 180^\circ ]$ phase range for $|\Gamma|=0$ dB. We observe that even in the case of non-ideal RIS-assisted transmission, there will be no significant changes in error performance. Another problem that may be encountered in real world applications is the imperfect estimation of the channel phase. This imperfect phase estimation results in the worsening of the received SNR by preventing the constructive combination of received signals. It is assumed that the phase estimation errors follow a zero-mean von Mises distribution with the concentration parameter $\kappa$ \cite{Coon_2019}, where $\kappa$ is a measure of the estimation accuracy. In Fig. \ref{non_ideal}(b), a single RIS-assisted system under imperfect and ideal phase estimation is investigated in terms of the error performance for varying $ \kappa $ and $ N $ values. Although small $\kappa$ values lead to a degradation in error performance, this effect is less noticeable for large $N$ and $\kappa$.

\section{Conclusions}
In this paper, we provide unique RIS-oriented solutions for potential communication scenarios that may emerge in 6G and beyond wireless networks. 
In this context, we investigate a number of RIS-assisted systems with single or multiple RISs, in terms of error performance, achievable data rate and path loss characteristics in indoor and outdoor environments.
To our knowledge, for the first time, our paper explores the effect of RISs on the overall path loss under empirical path loss models and the cases of multiple RISs, as well as RIS selection strategies, using a unified error performance framework. Our simulation results indicate that RIS-assisted link acts as an LOS path by suppressing the effect of blocked paths when the number of reflecting elements increases and double-RIS reflected transmission further improves the performance. In other words, one can eliminate the need for LOS transmission utilizing emerging RISs and multiple RIS-assisted systems may be a potential remedy for future dense wireless networks. 
We also note that the extension to MIMO and multi-user setups appears as an interesting future research direction. 
To sum up, the communication with real-time controlled RISs can be considered as an exciting technology to meet the demands of 6G and beyond wireless networks.

\bibliographystyle{IEEEtran}
\bibliography{IEEEabrv,bib_2019_eb}

\end{document}